\documentclass[reprint,amsmath,amssymb,aps,floatfix,prl,superscriptaddress]{revtex4-1}

\usepackage{graphicx}
\usepackage{dcolumn}
\usepackage{bm}

\begin{document}

\preprint{}

\title{Strong Coupling Effects in Binary Yukawa Systems}

\author{Gabor J. Kalman}
\affiliation{Department of Physics, Boston College, Chestnut Hill, MA 02467 USA}
\author{Zolt\'an Donk\'o}
\author{Peter Hartmann}
\affiliation{Research Institute for Solid State Physics and Optics of the Hungarian Academy of Sciences, P.O.B. 49, H-1525 Budapest, Hungary}
\affiliation{Department of Physics, Boston College, Chestnut Hill, MA 02467 USA}
\author{Kenneth I. Golden}
\affiliation{Department of Mathematics and Statistics, Department of Physics, University of Vermont, Burlington, VT 05401 USA}

\date{\today}

\begin{abstract}

We analyze the acoustic collective excitations in two- and three-dimensional binary Yukawa systems, consisting of two components with different masses. Theoretical analysis reveals a profound difference between the weakly and strongly correlated limits: at weak coupling the two components interact via the mean field only and the oscillation frequency is governed by the light component. In the strongly correlated limit the mode frequency is governed by the combined mass, where the heavy component dominates. Computer simulations in the full coupling range extend and confirm the theoretical results.

\end{abstract}


\maketitle

Recently there has been a great interest in the collective excitations of Yukawa liquids and solids, created by the emergence of the new field of complex (dusty) plasmas \cite{dusty}. Complex plasmas consist of highly charged mesoscopic grains immersed in the background of electrons and ions. It is the presence of the latter that, by screening the bare Coulomb interaction between the grains, generates an effective interaction that in a good approximation can be represented by the Debye-H\"uckel, or Yukawa potential $\phi(r)= Z e \exp(-\kappa r) /r$ ($\kappa$ is the screening parameter). The strength of the coupling governing the behavior of the systems is conventionally characterized by the nominal coupling constant $\Gamma=Z^2e^2/a k T$ ($a$ is the Wigner-Seitz radius and $T$ is the temperature). Due to the screening ($\kappa > 0$) the effective coupling constant $\Gamma^\ast$ (defined in \cite{PH,F}) may be substantially smaller. The high value of the grain charge ($Z \gg 1$) ensures that the system is in the strong coupling ($\Gamma^\ast \gg$ 1) regime and consequently in the liquid or solid phase. Both two-dimensional (2D) and three dimensional (3D) Yukawa systems (YS) are of interest, although most of the experimental work has focused so far on 2D systems. 

Over the years the collective excitations in YS-s have been studied experimentally, by computer simulations and theoretically. The analytic treatment of the crystalline solid phase is feasible via the harmonic phonon model. More challenging is the appropriate description of the liquid phase: here various approximation schemes have been attempted \cite{VE,MemF}, out of which the Quasilocalized Charge Approximation (QLCA) \cite{QLCA} has emerged with considerable success. All these efforts have by now congealed in a reasonably complete understanding of the collective mode spectra of liquid and solid YS-s, both in 2D and 3D. In addition, the YS has turned out to be a useful paradigm for other strongly coupled many body systems \cite{DP,R}.

The restriction in almost all of the foregoing investigations on collective excitations is that they address one component YS-s (OCYS), where all the particles carry the same charge number $Z$ and the same mass, $m$. Only a few exceptions are available, most notably the recent experimental realization of a two-component (binary) bilayer \cite{BBL}. Thus the collective mode structure of YS-s with more than one component, that of the \emph{two-component binary YS} (BYS) in particular, is still an open question. It is also by no means a trivial generalization of the single component problem: binary systems are well-known to exhibit a wealth of novel physical features: a much richer phase diagram, the degree of miscibility of different phases, new modalities of disorder, the excitation of optic modes are amongst them (e.g. \cite{Lowen,Jiang}). The connection to problems relating to other liquid or solid condensed matter systems are more immediate than in the single component case: the listing above provides a compass and as systems of expected interest binary ionic mixtures, ionic crystals, liquid and solid alloys and semiconductor bilayers come immediately to mind.

The asymmetry between the two components of a BYS is characterized by three parameters: the mass ratio $m_2/m_1$, the charge ratio $Z_2/Z_1$, and the density ratio $n_2/n_1$. In a complex plasma these parameters are not independent: most importantly, both the $m_2/m_1$ and the $Z_2/Z_1$ ratios are determined by the relative grain sizes. Theoretically and in simulation models, of course, these parameters can be separated; indeed they should be so distinguished, in order for one to be able to determine the different physical effects brought about by mass, charge, etc. asymmetries. We have already shown in \cite{1990} that for the purpose of calculating the dispersion relation the charge, mass and density ratios can be reduced to two asymmetry parameters $p^2 = (Z_2 n_2)/(Z_1 n_1)$ and $q^2 = (Z_2 m_1)/(Z_1 m_2)$.

This Letter addresses the issue of the collective spectrum of a BYS. We study the excitation of the longitudinal acoustic mode and we investigate how the asymmetry, the mass difference in particular, between the two species affects the sound speed as the coupling strength $\Gamma$  is varied from the weak coupling $\Gamma^\ast \ll 1 $ to the strong coupling $\Gamma^\ast \gg$1 regime. The model is analytically studied both for the weakly coupled and strongly coupled liquids and for the solid phase:  the first is done within the simple Random Phase Approximation (RPA) model, while for the  second we apply the QLCA, whose two-component version has already been worked out in earlier works on Coulomb systems \cite{1990}. For the solid phase we assume crystalline order and use the harmonic phonon approximation. For the liquid phase in the intermediate regime between the weak and strong coupling we follow the evolution of the collective modes as a function of the coupling strength through Molecular Dynamics (MD) simulation. These approaches yield a consistent picture: in a mixture of light and heavy components at weak coupling the two components interact through the mean field only and the oscillation frequency is governed by the light component. With increasing correlations, however, the dominance of the light component is diminished: in the strong coupling limit the light particles attach themselves to the heavy masses and the oscillation frequency is governed by the combined mass, where the heavy component dominates. As a result, one observes the remarkable and rather dramatic reduction of the sound velocity with increasing coupling.

Consider now a BYS, with masses and densities $m_A$ and $n_A$ ($A$=1,2), respectively. Each density can be associated with a Wigner-Seitz radius $a_A$ and a nominal coupling constant $\Gamma_A=Z_A^2e^2/a_A k T$. The intraction potentials are $\varphi_A^{\rm 3D}(k) = 4 \pi Z_A^2 e^2/(k^2 + \kappa^2)$ and $\varphi_A^{\rm 2D}(k) = 2 \pi Z_A^2 e^2/\sqrt{k^2 + \kappa^2}$. In a OCYS the longitudinal collective mode is acoustic at long wavelength ($k \rightarrow 0$), and has a coupling dependent sound velocity $s$ \cite{RK,QLCA}. It is the equivalent of this mode in the BYS that we concentrate on. In the calculations we parallel the 2D and 3D results. In the weak coupling limit the sound velocity is determined by the dispersion relation obtained from the RPA dielectric function: 
\begin{equation}
\varepsilon_{\rm RPA} = 1 - \sum_A \varphi_A(k) \chi_A^0(k,\omega) ~; ~~~
\chi_A^0 (k,\omega) = \frac{n_A}{m_A} \frac{k^2}{\omega^2}.
\label{eq:eq1}
\end{equation}
For our purpose it is sufficient to use the $T$=0 ``cold fluid'' result: temperature dependent terms slightly increase the sound speed over this value. Eq. (\ref{eq:eq1}) yields the longitudinal sound velocity $s$:
\begin{eqnarray}
s^{\rm 3D} = \omega_0 / \kappa ~~~~{\rm and}~~~~ s^{\rm 2D} = \omega_0 \sqrt{a/ \kappa}~~;~~~~~~~~~~~~\\ \label{eq:eq2}
\omega_A^{\rm 3D}=\sqrt{4 \pi Z_A^2 e^2 n_A/m_A},~~~ \omega_A^{\rm 2D}=\sqrt{2 \pi Z_A^2 e^2 n_A / m_A a}. 
\nonumber 
\end{eqnarray}
$\omega_0= \sqrt{\omega_1^2 + \omega_2^2} = \omega_1 \sqrt{1+p^2 q^2}$ is the total plasma frequency and $a$=$\sqrt{a_1 a_2}$. The salient feature of the RPA result is that through the total plasma frequency the sound speed is governed by the reduced mass. i.e. by the \emph{light} component and, with a sizable difference between the light and heavy masses, the presence of the heavy component plays a negligible role.

In the strong coupling limit we calculate the dispersion using the QLCA with the input of correlation functions obtained from MD simulations. The QLCA is expected to provide a reliable description of the strongly coupled liquid system, as it is attested by its application both to 2D and to 3D OCYL-s. (see, e.g. \cite{QLCA,PRL}. While we do not know how faithfully the QLCA can describe the mode structure in binary systems, there is little doubt that it is reliable in the long-wavelength ($k \rightarrow 0$) limit. This is evidenced by the demonstration \cite{IEEE} that it provides a smooth transition to the angle-averaged long-wavelength phonon dispersion of a corresponding crystal lattice.  

In the QLCA the central quantity is the  longitudinal dynamical matrix, $C_{AB}^{L}(\mathbf{k})$ dependent on the  equilibrium pair correlation function $h_{AB}(r)$. We note that since masses do not affect equilibrium quantities, with appropriate scaling, all the $h_{AB}$-s are identical and derivable from $h_0(r)$, the pair correlation function for the single component system. The dispersion relation expressed in terms $C_{AB}^{L}(\mathbf{k})$ follows from
\begin{equation}
|| \omega^2 \delta_{AB} - C_{AB}^{L}(k) || = 0.
\label{eq:eq4}
\end{equation}		
We consider first the 3D YBS.  
In the long-wavelength ($k \rightarrow 0$) limit of interest, the QLCA matrix elements are 
\begin{eqnarray}
C_{11}^{L}(\mathbf{k} \rightarrow 0) = \omega_{1}^2 \biggl[ (1-U_{11}^{\rm 3D}) \frac{k^2}{\kappa^2} + \frac{p^2}{3} W 
\biggr] \nonumber \\
C_{12}^{L}(\mathbf{k} \rightarrow 0) = \omega_{1}^2 \biggl[ p q(1-U_{12}^{\rm 3D}) \frac{k^2}{\kappa^2} - \frac{pq}{3} W 
\biggr] \nonumber \\
C_{22}^{L}(\mathbf{k} \rightarrow 0) = \omega_{1}^2 \biggl[ p^2 q^2 (1-U_{22}^{\rm 3D}) \frac{k^2}{\kappa^2} + \frac{q^2}{3} W \biggr] \\
\label{eq:eq4}
U_{AB}^{\rm 3D} = - \frac{2}{15} \int_0^\infty {\rm d}y ~y \biggl[ 1+y + \frac{3}{4}y^2 \biggr] {\rm e}^{-y} h_{AB}(r), 
\nonumber \\
W = 1 + \int_0^\infty {\rm d}y~ y {\rm e}^{-y} h_{12}(r);~~~ y = \kappa r. \nonumber
\end{eqnarray}
Introducing the average charge and mass, 
\begin{equation}
\langle Z \rangle = \frac{Z_1 n_1 + Z_2 n_2}{n_1 + n_2}, \nonumber ~~~
\langle m \rangle = \frac{m_1 n_1 + m_2 n_2}{n_1 + n_2}, 
\label{eq:eqxx}
\end{equation}		
we find that the crucial frequency parameter now is
\begin{equation}
\widetilde{\omega}^{\rm 3D} = \sqrt{4 \pi n_0 e^2 \frac{\langle Z \rangle^2 }{\langle m \rangle }} =
\omega_1^{\rm 3D} ~\frac{q (1+p^2)}{\sqrt{p^2+q^2}},
\label{eq:eq7}
\end{equation}
a quantity that has been dubbed in the literature as the frequency related to ``the pseudo-alloy atom'' (FPAA) \cite{Vora}, or to the average atom in the virtual crystal approximation \cite{Fultz,Poon,Langer}. From Eqs. (3) and (4) one now readily obtains the 3D sound speed as
\begin{equation}
s^{\rm 3D} = \frac {\widetilde{\omega}^{\rm 3D}} {\kappa} \sqrt{1-U^{\rm 3D}(\Gamma)}.
\label{eq:s3d}
\end{equation}
A similar calculation in 2D leads to:
\begin{eqnarray}
\widetilde{\omega}^{\rm 2D} = \sqrt{2 \pi n_0 e^2 \frac{\langle Z \rangle^2 }{\langle m \rangle a}}, ~~~~
s^{\rm 2D} =\widetilde{\omega}^{\rm 2D} \sqrt{\frac{a}{\kappa}} \sqrt{1-U^{\rm 2D}(\Gamma)};
\nonumber \\
U_{AB}^{\rm 2D} = -\frac{5}{16} \int_0^\infty {\rm d}y \Bigl[ 1 + y + \frac{3}{5} y^2 \Bigr] {\rm e}^{-y} h_{AB} (r).~~~~~~~
\label{eq:s2d}
\end{eqnarray}
$U^{\rm 3D}$ and $U^{\rm 2D}$ in Eqs.(\ref{eq:s3d}) and (\ref{eq:s2d}), respectively, are calculated from $U = \bigl[ U_{11} + 2 p^2 U_{12} + p^4 U_{22} \bigr] / (1+p^2)^2$, using the corresponding $U_{AB}^{\rm 3D}$ and $U_{AB}^{\rm 2D}$ values. Note that for $q=1$ we have $\widetilde{\omega}=\omega_0$ and one recovers a quasi-single component behavior \cite{1990}. 

There are two important issues to be noted here. First, $\widetilde{\omega}$, in contrast to $\omega_0$, is governed by the \emph{heavier} component. Second, the leading terms in the equations for  $s^{\rm 3D}$ and $s^{\rm 2D}$ are seemingly coupling independent, although these equations are valid in the strong coupling limit, since the QLCA is a strong coupling approximation. The explicit coupling dependence enters through the $U_{AB}(\Gamma)$ terms only.

Currently, there is no theoretical understanding of the transition bridging the weak, light species dominated, and the strong, heavy species dominated, coupling domains. In order to follow the collective excitations in this intermediate coupling region and to verify the predictions of the QLCA theory, we generate 3D and 2D MD simulation data for the longitudinal current-current dynamical structure factors, from which we obtain the acoustic speed, over a wide range of $\Gamma$ values extending well into the crystalline solid region. We analyze the liquid phase of several systems, described below, in the $5 \leq \Gamma_1 \leq \Gamma_{\rm m}$ coupling domain; $\Gamma_{\rm m}$ is the coupling value where the system solidifies (melts). The lower limit of $\Gamma$ is set by the limitations of the simulation technique.

\begin{figure}[htb]
\includegraphics[width=\columnwidth]{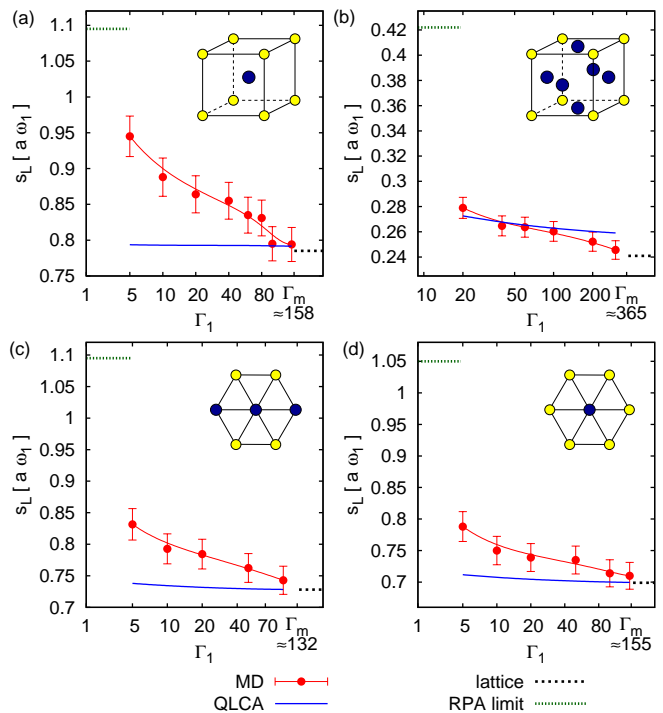}
\caption{\label{fig:1} 
(color online) Longitudinal sound speed vs. $\Gamma_1$ in different systems, obtained from MD simulations. 3D: (a) $n_2=n_1$, $\kappa$ = 1, (b) $n_2/n_1=3$, $\kappa=3$. 2D: (c) $n_2=n_1$, $\kappa$=1, (d) $n_2/n_1 = 0.5$, $\kappa$=1. $m_2 / m_1$=5 for all cases. The dotted lines indicate the theoretical low $\Gamma$ and high $\Gamma$ limits; the continuous lines represents the results of the QLCA calculations. The panels also illustrate the respective crystal structures in the solid phases. }
\end{figure}

While the theoretical results are quite general, covering the case of a binary Yukawa system for any asymmetry, the simulations are confined to systems with $m_1 \neq m_2$ but $Z_1 = Z_2$. We study both 2D and 3D systems, both with equal densities ($n_1=n_2$) and with unequal densities: for the latter we have chosen $n_2=(1/2) n_1$ in 2D and $n_2= 3 n_1$ in 3D.

At $\Gamma \geq \Gamma_{\rm m}$, the system is in the solid phase, in 2D the underlying lattice structure is hexagonal; in 3D it is bcc or fcc, depending on the value of $\kappa$ \cite{Hamaguchi}. With mass-asymmetry only, the occupation of the lattice sites by the two species would be random, since the potential energy is independent of how particles are distributed, and entropy prefers the disordered distribution. It is only when in addition to unequal masses ($m_1 \neq m_2$), the charges are also unequal ($Z_1 \neq Z_2$), that a particular crystal structure is generated over the lattice in order to minimize the energy. Nevertheless, in this work, in anticipation to its relevance to realistic systems, we have studied crystal structures with maximally symmetric distribution of the two components. The corresponding symmetric crystal structures are shown in the insets in Fig.~1. (We have also verified by simulation that the chosen structures would be stable configurations for the Yukawa solids, should the $m_2/m_1$ ratio be accompanied by $Z_2/Z_1 \cong \sqrt[3]{m_2/m_1}$, as it happens in dusty plasmas.) The sound velocities for these crystal structures were calculated in the harmonic phonon approximation; these were also confirmed by MD simulations (of systems with $\Gamma \sim 10^4$, initiated with lattice configurations). These values and sound velocities obtained from simulations for the disordered phase are virtually indistinguishable.

Our main results for the sound speed are portrayed in Fig. 1, where the four panels correspond to the four cases listed above. In addition to the results of the MD simulations, the QLCA prediction for the high $\Gamma$ liquid, and the results pertaining to the long wavelength phonons in the respective crystal structures. For the weak coupling domain we have indicated a representative value, the ``cold fluid'' sound velocity [Eq.(1)]: inclusion of thermal effects would  only slightly increase this value. Fig. 1 shows the rather dramatic decrease of the sound velocity with increasing $\Gamma$ from the weakly coupled to the strongly coupled regions. For high $\Gamma$, near $\Gamma_{\rm m}$  the agreement of the MD results with the QLCA predictions is excellent. There is also an almost perfect agreement near $\Gamma_{\rm m}$ between these two values pertaining to the liquid and the crystal lattice values. The small discrepancy, visible for the 3D $n_2= 3 n_1$ case can be attributed to the anisotropy of the sound velocity in the fcc lattice: the liquid results correspond to angle-averaged values, while the lattice result is given along the chosen \{001\} direction. This point is further elucidated in Fig.~2, where the dependence of the sound velocities on the mass ratio is shown for the different crystalline solid phases and are compared with the QLCA predictions.

Our calculations show (Fig. 1) that the $\Gamma$-dependence of the sound speed is weak and that even at the crystallization boundary it does not amount to more than 10\% of the leading term. In view of our earlier statement, it is now safe to assume that the validity of the QLCA result can be extended to the crystal lattice region. The importance of this observation lies in the fact that the notion of FPAA has been successfully used for the description of phonon dispersion for binary alloys \cite{Vora,Fultz} and disordered systems \cite{Poon,Langer}, as a heuristic concept. Here we have provided an analytic derivation of this behavior.
 
\begin{figure}[htb]
\includegraphics[width=0.8\columnwidth]{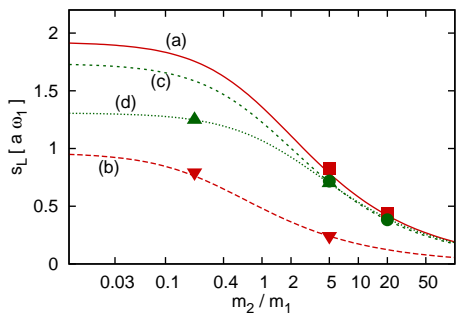}
\caption{\label{fig:2} 
(color online) Longitudinal sound speed vs. $m_2/m_1$, obtained from lattice calculations. The labels correspond to the  cases given in Fig 1. The  symbols represent QLCA calculation results for the strongly coupled liquid phase.}
\end{figure}
  
\begin{figure}[htb]
\includegraphics[width=0.8\columnwidth]{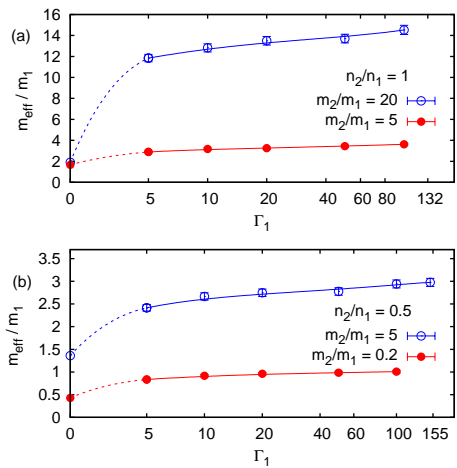}
\caption{\label{fig:3} 
(color online) Effective mass vs. $\Gamma_1$, obtained from MD simulations, for 2D sytems at different mass ratios and densities, at $\kappa$=1. The dotted lines represent extrapolation to low $\Gamma$ RPA values.}
\end{figure}
 
The physical effect that causes the lowering of the sound velocity with increasing correlations may be attributed to the binding of the lighter particles to the heavier ones, resulting in a combined effective mass. This latter, defined (e.g. in 2D) by
\begin{equation}
\frac{m_{\rm eff}}{m_1} = \frac{n_1+n_2}{n_1} \frac{1}{\kappa a} \frac{1}{(s^{\rm 2D}/\omega_1 a)^2},
\end{equation}
is displayed in Fig. 3, as a function of the coupling. It may be noted that the binding seems to become quite pronounced already at the relatively low $\Gamma$=5 value. 

In summary, we have analyzed the behavior of the acoustic excitation in a binary Yukawa system consisting of two components with different masses, as a function of the plasma coupling strength $\Gamma$. Our main focus has been to see how correlations affect the way the two masses bind into an effective mass. Theoretical analysis at the weakly and strongly correlated limits shows that while in the weakly correlated ($\Gamma < 1$) system the effective mass forms as the reduced mass (``parallel connection''), in the strongly correlated liquid or solid phase ($\Gamma \gg 1$) the effective mass is simply the weighted average of the two masses (``series connection''). As a result, the sound speed is substantially diminished in the strong coupling domain, as compared to its weak coupling value. Our MD simulations of the longitudinal acoustic mode, straddling the intermediate coupling range, have confirmed the result and have mapped the variation of the sound speed over a wide range of $\Gamma$ values, both for 3D and 2D systems.

A remarkable feature of the derivation is that the leading term in the strong coupling expression is formally correlation independent: it is a consequence of the localization of the particles, inherent in the model. Thus, even though our analysis pertains only to systems interacting through a Yukawa potential, one may expect that it has a more general validity and other binary systems with an acoustic or quasi-acoustic type excitation (a two-dimensional Coulomb liquid in particular) would follow a similar pattern. 

From the experimental point of view, we note that in alloys -- bearing a great deal of similarity to Yukawa solids -- the notion of the ``pseudo-alloy atom'' is heuristically well established (see e.g. \cite{Vora} and references therein). As to complex plasmas, laboratory experiments on binary systems could be more feasible in 3D than in 2D, where they should require levitating grains with two different $Z/m$ ratios.  

\vspace{0.5cm}

\begin{acknowledgments}
The authors appreciate helpful inputs from Stamatios Kyrkos and Marlene Rosenberg. This work has been partially supported by NSF Grants PHY-0813153, PHY-0715227, PHY-0812956, OTKA Grants K-77653 and PD-75113, and the J\'anos Bolyai Research Foundation of the Hungarian Academy of Sciences.
\end{acknowledgments}

\end{document}